


\input lanlmac

\input epsf

\newcount\figno
\figno=0
\def\fig#1#2#3{
\par\begingroup\parindent=0pt\leftskip=1cm\rightskip=1cm\parindent=0pt
\baselineskip=11pt
\global\advance\figno by 1
\midinsert
\epsfxsize=#3
\centerline{\epsfbox{#2}}
\vskip 12pt
\centerline{{\bf Fig. \the\figno:~~} #1}\par
\endinsert\endgroup\par
}
\def\figlabel#1{\xdef#1{\the\figno}}


\def\cT{{\cal T}}

\def\Z{{\bf Z}}

\def\hf{{1\over 2}}

\def\R{{\bf R}}
\def\o{\over}

\def\del{\partial}
\def\wg{\wedge}
\def\lap{\Delta}
\def\bra{\langle}
\def\ket{\rangle}
\def\lf{\left}
\def\ri{\right}
\def\riya{\rightarrow}

\def\la{\lambda}

\def\Ga{\Gamma}
\def\al{\alpha}

\def\rt#1{\sqrt{#1}}

\def\sitarel#1#2{\mathrel{\mathop{\kern0pt #1}\limits_{#2}}}

\def\cob{\delta}

\lref\SenMG{
A.~Sen,
``Non-BPS states and branes in string theory,''
[hep-th/9904207].
}

\lref\Kutasov{
D. Kutasov and V. Niarchos,
``Tachyon Effective Actions In Open String Theory,''
[hep-th/0304045].
}

\lref\Sbrane{
M.~Gutperle and A.~Strominger,
``Spacelike branes,''
JHEP {\bf 0204}, 018 (2002)
[hep-th/0202210].
}

\lref\Drukker{
N.~Drukker, D.~J.~Gross and N.~Itzhaki,
``Sphalerons, merons and unstable branes in AdS,''
Phys.\ Rev.\ D {\bf 62}, 086007 (2000)
[hep-th/0004131].
}

\lref\SenNU{
A.~Sen,
``Rolling tachyon,''
JHEP {\bf 0204}, 048 (2002)
[hep-th/0203211].
}

\lref\SenTM{
A.~Sen,
``Tachyon matter,''
JHEP {\bf 0207}, 065 (2002)
[hep-th/0203265].
}

\lref\SenT{
A.~Sen,
``Field theory of tachyon matter,''
Mod.\ Phys.\ Lett.\ A {\bf 17}, 1797 (2002)
[hep-th/0204143].
}

\lref\SenTime{
A.~Sen,
``Time and tachyon,''
[hep-th/0209122].
}

\lref\SenBI{
A.~Sen,
``Dirac-Born-Infeld action on the tachyon kink and vortex,''
[hep-th/0303057].
}

\lref\Sthermo{
A.~Maloney, A.~Strominger and X.~Yin,
``S-brane thermodynamics,''
[hep-th/0302146].
}

\lref\Lambert{
N.~D.~Lambert and I.~Sachs,
``Tachyon dynamics and the effective action approximation,''
Phys.\ Rev.\ D {\bf 67}, 026005 (2003)
[hep-th/0208217].
}

\lref\Spha{
J.~A.~Harvey, P.~Horava and P.~Kraus,
``D-sphalerons and the topology of string configuration space,''
JHEP {\bf 0003}, 021 (2000)
[hep-th/0001143].
}
\lref\LLM{
N.~Lambert, H.~Liu and J.~Maldacena,
``Closed strings from decaying D-branes,''
[hep-th/0303139].
}

\lref\KrausBSFT{
P.~Kraus and F.~Larsen,
``Boundary string field theory of the DD-bar system,''
Phys.\ Rev.\ D {\bf 63}, 106004 (2001)
[hep-th/0012198].
}

\lref\TakayaBSFT{
T.~Takayanagi, S.~Terashima and T.~Uesugi,
``Brane-antibrane action from boundary string field theory,''
JHEP {\bf 0103}, 019 (2001)
[hep-th/0012210].
}

\lref\SenVV{
A.~Sen,
``Time evolution in open string theory,''
JHEP {\bf 0210}, 003 (2002)
[hep-th/0207105].
}

\lref\MukhopadhyayEN{
P.~Mukhopadhyay and A.~Sen,
``Decay of unstable D-branes with electric field,''
JHEP {\bf 0211}, 047 (2002)
[hep-th/0208142].
}
\lref\StromingerPC{
A.~Strominger,
``Open string creation by S-branes,''
[hep-th/0209090].
}

\lref\ChenFP{
B.~Chen, M.~Li and F.~L.~Lin,
``Gravitational radiation of rolling tachyon,''
JHEP {\bf 0211}, 050 (2002)
[hep-th/0209222].
}

\lref\Cline{
J.~M.~Cline and H.~Firouzjahi,
``Real-time D-brane condensation,''
[hep-th/0301101].
}

\lref\ReyXS{
S.~J.~Rey and S.~Sugimoto,
``Rolling tachyon with electric and magnetic fields: T-duality approach,''
[hep-th/0301049].
}

\lref\MoellerVX{
N.~Moeller and B.~Zwiebach,
``Dynamics with infinitely many time derivatives and rolling tachyons,''
JHEP {\bf 0210}, 034 (2002)
[hep-th/0207107].
}

\lref\SugimotoFP{
S.~Sugimoto and S.~Terashima,
``Tachyon matter in boundary string field theory,''
JHEP {\bf 0207}, 025 (2002)
[hep-th/0205085].
}

\lref\MinahanIF{
J.~A.~Minahan,
``Rolling the tachyon in super BSFT,''
JHEP {\bf 0207}, 030 (2002)
[hep-th/0205098].
}

\lref\KlusonAV{
J.~Kluson,
``Exact solutions in open bosonic string field theory and marginal  deformation in CFT,''
[hep-th/0209255].
}

\lref\KlusonTE{
J.~Kluson,
``Time dependent solution in open bosonic string field theory,''
[hep-th/0208028].
}
\lref\Larsen{
F.~Larsen, A.~Naqvi and S.~Terashima,
``Rolling tachyons and decaying branes,''
JHEP {\bf 0302}, 039 (2003)
[hep-th/0212248].
}

\lref\OkudaYD{
T.~Okuda and S.~Sugimoto,
``Coupling of rolling tachyon to closed strings,''
Nucl.\ Phys.\ B {\bf 647}, 101 (2002)
[hep-th/0208196].
}

\lref\ReyZJ{
S.~J.~Rey and S.~Sugimoto,
``Rolling of modulated tachyon with gauge flux and emergent fundamental  string,''
[hep-th/0303133]
}

\lref\ArefevaQU{
I.~Y.~Aref'eva, L.~V.~Joukovskaya and A.~S.~Koshelev,
``Time evolution in superstring field theory on non-BPS brane. I: Rolling  tachyon and energy-momentum conservation,''
[hep-th/0301137].
}

\lref\IshidaCJ{
A.~Ishida and S.~Uehara,
``Rolling down to D-brane and tachyon matter,''
JHEP {\bf 0302}, 050 (2003)
[hep-th/0301179].
}

\lref\Gutperle{
M.~Gutperle and A.~Strominger,
``Timelike Boundary Liouville Theory,''
[hep-th/0301038].
}

\lref\Peet{
F.~Leblond and A.~W.~Peet,
``SD-brane gravity fields and rolling tachyons,''
[hep-th/0303035].
}

\lref\SenMD{
A.~Sen,
``Supersymmetric world-volume action for non-BPS D-branes,''
JHEP {\bf 9910}, 008 (1999)
[hep-th/9909062].
}

\lref\GarousiTR{
M.~R.~Garousi,
``Tachyon couplings on non-BPS D-branes and Dirac-Born-Infeld action,''
Nucl.\ Phys.\ B {\bf 584}, 284 (2000)
[hep-th/0003122].
}
\lref\GarousiWQ{
M.~R.~Garousi,
``On-shell S-matrix and tachyonic effective actions,''
Nucl.\ Phys.\ B {\bf 647}, 117 (2002)
[hep-th/0209068].
}

\lref\BergshoeffDQ{
E.~A.~Bergshoeff, M.~de Roo, T.~C.~de Wit, E.~Eyras and S.~Panda,
``T-duality and actions for non-BPS D-branes,''
JHEP {\bf 0005}, 009 (2000)
[hep-th/0003221].
}

\lref\KlusonIY{
J.~Kluson,
``Proposal for non-BPS D-brane action,''
Phys.\ Rev.\ D {\bf 62}, 126003 (2000)
[hep-th/0004106].
}

\Title{             
                                             \vbox{\hbox{EFI-03-15}
                                             \hbox{hep-th/0304108}}}
{\vbox{
\centerline{Wess-Zumino Term in Tachyon Effective Action}
}}

\vskip .2in

\centerline{Kazumi Okuyama}

\vskip .2in

\centerline{ Enrico Fermi Institute, University of Chicago} 
\centerline{ 5640 S. Ellis Ave., Chicago IL 60637, USA}
\centerline{\tt kazumi@theory.uchicago.edu}

\vskip 3cm
\noindent


We show that the source of RR field computed from the boundary state 
describing the decay of a non-BPS brane is reproduced
by a particular form of the Wess-Zumino term in the tachyon
effective action. 
We also obtain a simple expression of the
S-charge associated with rolling tachyons.

\Date{April 2003}

\vfill
\vfill

\newsec{Introduction}
Decay of an unstable brane is a 
very interesting process which might help us to understand
some properties of string theory in time-dependent backgrounds
\refs{\Sbrane\SenNU\SenTM\SenT\SenVV\SenTime\MukhopadhyayEN\StromingerPC\Larsen\Gutperle\Sthermo\OkudaYD\ReyXS\ReyZJ{--}\LLM}. 
The dynamics of tachyon field $\cT$ on such a brane
can be described by an effective field theory
when the value of $\cT$ and its derivatives satisfy
some conditions \SenTime. 
The proposed effective action of real tachyon $\cT$ on non-BPS $Dp$-brane
in Type II string theory 
is given by \refs{\SenMD\GarousiTR\BergshoeffDQ\KlusonIY{--}\GarousiWQ,\SenT}
\eqn\VTac{
S_{DBI}=\int d^{p+1}x{\cal L}=-M_p\int d^{p+1}x
V(\cT)\rt{1+\del_\mu\cT \del^\mu \cT}
}
where $M_p$ is the tension of the non-BPS brane.
In this paper we use the unit $\al'=1$.
In \VTac, we suppressed the dependence on the 
gauge field and the scalar fields other than the tachyon.
It was pointed out in \LLM\ that
if we take the potential $V(\cT)$ to be
\eqn\VT{
V(\cT)={1\o\cosh{\cT\o\rt{2}}},
}
then the stress tensor of rolling tachyon 
computed from the boundary state
is correctly reproduced by the effective action \VTac.
This potential \VT\ was also discussed recently in \refs{\Peet,\Kutasov}.

The tachyon effective action has another important term, {\it i.e.} 
the Wess-Zumino term (or the Chern-Simons term) describing 
the coupling of open string tachyon to the bulk RR field:
\eqn\SCR{
S_{WZ}=\int W(\cT)d\cT\wg C_{RR}
}
where $W(\cT)$ is an even function of $\cT$ which vanishes as
$\cT\riya\pm\infty$.
In order 
for a kink of $\cT$ connecting the two vacua
at $\cT=\pm\infty$ to carry the correct RR charge and the tension
of BPS $D(p-1)$ brane, 
$W(\cT)$ and $V(\cT)$ should satisfy the condition \SenBI 
\foot{Our definition of $W(\cT)$ 
is different from that in \SenBI\
by the factor of $M_{p-1}^{BPS}$.}
\eqn\WvsVconsis{
\int_{-\infty}^\infty W(\cT)d\cT
={M_p\o M_{p-1}^{BPS}}\int_{-\infty}^\infty V(\cT)d\cT
=1.
}
Here $M_{p-1}^{BPS}$ is the tension of BPS $D(p-1)$-brane whose 
ratio to the tension $M_p$ of non-BPS $Dp$-brane is given by \SenMG
\eqn\MptoMpp{
{M_p\o M_{p-1}^{BPS}}={1\o\rt{2}\pi}.
}
Note that the last equality in \WvsVconsis\
is satisfied for $V(\cT)={1\o\cosh({\cT\o\rt{2}})}$.

In this short note, we will show that the coupling of the open string 
tachyon to the RR field
computed from the boundary state is reproduced by setting
\eqn\WvsV{
W(\cT)={1\o\rt{2}\pi}V(\cT)={1\o\rt{2}\pi\cosh{\lf({\cT\o\rt{2}}\ri)}}.
}
This $W(\cT)$ obviously satisfies the condition \WvsVconsis.\foot{
From the late time behavior of the source 
for the RR field in the boundary state,
Sen showed that $W(\cT)$ goes like $e^{-{1\o\rt{2}}\cT}$
for large $\cT$ \SenT. 
\WvsV\ has this property.
}

In section 2, we consider some properties
of the equation of motion obtained from
the action \VTac\ with $V(\cT)$ given by \VT.
In section 3, we show that the source of RR field computed 
from the boundary state
is reproduced by \WvsV. We also comment on the 
S-charge associated with the rolling tachyon.
We conclude with a few comments in section 4.

\newsec{Some Properties of $S_{DBI}$}
To see the dynamics of tachyon field described by the action \VTac\ with
\VT, it is convenient to perform a field redefinition from $\cT$
to $T$ as \Kutasov
\eqn\ctvsT{
\sinh{\cT\o\rt{2}}={T\o\rt{2}}.
}
In terms of this variable $T$, the Lagrangian becomes
\eqn\LinT{
{{\cal L}\o M_p}=-{1\o 1+\hf{T^2}}\rt{1+\hf{T^2}+\del_\mu T\del^\mu T},
}
and the equation of motion derived from this Lagrangian is
\eqn\eomT{
\lf(\del_\mu \del^\mu T+\hf T\ri)\lf(1+\hf{T^2}\ri)
+\del_\mu T\del^\mu T\del_\nu\del^\nu T-
\del_\mu T\del^\nu T\del_\nu\del^\mu T=0.
}
When the tachyon $T$ depends only on time $t$,
the second and the third terms in \eomT\ cancel, and the 
equation of motion reduces to
\eqn\Teqhom{
-\del_t^2T+\hf T=0.
} 
Therefore, the general solution of homogeneous tachyon is given by\foot{
The converse statement was proved in \Kutasov.
They showed that the effective action is fixed to have the form
\LinT\
by requiring: (1) $T=e^{{1\o\rt{2}}t}$ is a solution of 
the equation of motion at each order in the power series expansion
in $T$, (2) The on-shell Lagrangian agrees with the
disk partition function.
}
\eqn\Thom{
T=T_+e^{{1\o\rt{2}}t}+T_-e^{-{1\o\rt{2}}t}
}
with some constant coefficient $T_\pm$.

An interesting fact is that the form of
stress tensor as a function of time 
is reproduced from this effective action \LLM.
For instance, for the tachyon configuration describing
the ``full brane''
\eqn\fullT{
{T\o\rt{2}}={\cosh{{t\o\rt{2}}}\o\sinh{{t_0\o\rt{2}}}},
}
the on-shell Lagrangian is evaluated as
\eqn\onL{
-{{\cal L}\o M_p}={1\o 1+e^{\rt{2}(t-t_0)}}
+{1\o 1+e^{-\rt{2}(t+t_0)}}-1.
}
This expression agrees with the disk partition function $Z(x^0)$ 
in the presence of boundary tachyon operator 
$T_{WS}=\la\cosh (X^0/\rt{2})$,  
if we identify the coupling $\la$ 
and the parameter $t_0$ in \fullT\ as
\eqn\latotn{
\sin\pi\la=e^{-{1\o\rt{2}}t_0}.
}

In general, the worldsheet tachyon operator $T_{WS}$ in the $(-1)$ picture
is different from the tachyon field $T$ in the effective action
\LinT. In particular, the periodicity of 
$\la$ cannot be reproduced by any first derivative effective actions \Lambert.
Moreover,  the relative coefficient
of $T_{00}$ and $T_{ij}$ computed from \VTac\ does not 
agree with the boundary state result for the full brane.
As pointed out in \SenTime, if a tachyon configuration has 
a turning point $\del_0T=0$, the effective action 
\VTac\ cannot reproduce the stress tensor computed from the
boundary state.

However, after taking the ``half S-brane'' limit,
\eqn\halflim{
t\riya t+a,\quad t_0\riya t_0+a,\quad a\riya\infty
\quad{\rm with}~~t,t_0~~{\rm fixed},
}
we can actually identify $T_{WS}$ with $T$.
In this limit, 
the tachyon configuration and the on-shell Lagrangian become
\eqn\TandLlim{
{T\o\rt{2}}=e^{{1\o\rt{2}}(t-t_0)},\quad
-{{\cal L}\o M_p}={1\o 1+e^{\rt{2}(t-t_0)}},
}
and the stress tensor
agrees with the worldsheet computation for the half S-brane
with $T_{WS}=\la e^{{1\o\rt{2}}X^0}$ \refs{\Larsen,\Gutperle}.
It was argued in \Kutasov\
that the action \VTac\ is reliable only for the fluctuation 
of tachyon field around
a single exponential $e^{\pm t/\rt{2}}$
({\it i.e.} $T_+=0$ or $T_-=0$ in \Thom).

Next let us briefly
discuss the inhomogeneous tachyon configuration.
It is easy to see that the configuration with
a single exponential
\eqn\singleE{
T=e^{ik_\mu X^\mu}
}
solves the equation of motion \eomT\ provided
$k_\mu$ is on-shell:
\eqn\onshell{
k_\mu k^\mu=\hf.
}
However, the sum of on-shell exponentials, {\it e.g.}
\eqn\ecosT{
T=\hf e^{{1\o 2}(t+ix)}+\hf e^{\hf(t-ix)}
=e^{\hf t}\cos\Big(\hf x\Big)
}
is not a solution of \eomT\ because of the
non-linearity of the equation of motion.
It is interesting that this effective action \LinT\
knows that the boundary interaction $e^{X^0/2}\cos(X/2)$ is not 
exactly marginal in the superstring case \SenVV, 
although the single exponential $e^{(X^0\pm iX)/2}$ is exactly marginal. 
It seems  difficult to find an exact inhomogeneous
solution of \LinT. Instead, we can solve \LinT\
numerically, or solve it perturbatively along the line 
of \SenVV.
It was observed that, during the inhomogeneous decay of an unstable brane,
defect branes are formed at a finite time \refs{\SenVV,\Larsen}
and the slope of the kink becomes
infinitely steep \Cline. 
It would be interesting to see whether the tension of 
the defect brane calculated by this effective action 
agrees with the tension of the BPS brane.

\newsec{Wess-Zumino Term and S-charge}
In this section we will show that the 
source of RR field read off from the boundary state 
is reproduced by the Wess-Zumino term \SCR\ with the choice 
$W(\cT)={1\o\rt{2}\pi} V(\cT)$.

The source of RR field generated by the rolling tachyon 
$T_{WS}=\la\cosh X^0/\rt{2}$ is \SenT
\eqn\JRR{\eqalign{
j(t)&={\sin\pi\la\o\rt{2}\pi}
\lf[{e^{{1\o\rt{2}}t}\o1+\sin^2\pi\la\, e^{\rt{2}t}}
-{e^{-{1\o\rt{2}}t}\o1+\sin^2\pi\la\, e^{-\rt{2}t}}\ri].
}}
This source and the bulk RR $p$-form 
field couple via the Wess-Zumino term on the
worldvolume of non-BPS $Dp$-brane
\eqn\RRcouple{
S_{WZ}=\int_{\R^{1,p}} J\wg C_{RR}=\int dt d^p\xi\,j(t)C_{1,\cdots,p}
}
where $J=j(t)dt$.

Using the parameter $t_0$ \latotn, the source $j(t)$ is rewritten as 
\eqn\Jtzero{
j(t)={1\o\rt{2}\pi}\lf[{e^{{1\o\rt{2}}(t-t_0)}\o1+e^{\rt{2}(t-t_0)}}
-{e^{-{1\o\rt{2}}(t+t_0)}\o1+e^{-\rt{2}(t+t_0)}}\ri] 
={1\o\rt{2}\pi}\cdot{\sinh {t\o\rt{2}}\sinh{t_0\o\rt{2}}\o
\cosh^2{t\o\rt{2}}+\sinh^2{t_0\o\rt{2}}}.
}
Note that the normalization of $j(t)$ was determined 
by requiring that the S-brane with $T_{WS}=\la\sinh X^0/\rt{2}$
should carry $\pm 1$ of unit RR charge \Sthermo.
$j(t)$ has poles at
$t=\pm t_0+\pi i(2n+1)/\rt{2}$ and the residues are  
\eqn\Resj{
2\pi i \,{\rm Res}\,j(t)\Big|_{t=\pm t_0+{\pi i\o \rt{2}}(2n+1)}
=\pm (-1)^n\qquad (n\in \Z).
}
The alternating sign of residues corresponds to the
fact that, when $t_0=0$ (or $\la=\hf$), the Wick rotated 
configuration describes 
an array of branes and anti-branes.

From 
\Jtzero, \fullT\ and \ctvsT, one can easily see that
the RR source $J$ is written as
\eqn\JRRinT{
J={1\o\rt{2}\pi}{dT\o 1+\hf T^2}={1\o\rt{2}\pi}V(\cT)d\cT.
}
This relation holds also for the half S-brane. (The RR source of 
the half S-brane is obtained by taking the limit \halflim\ in \JRR.)
Therefore, we conclude that the function $W(\cT)$ is given by
${1\o\rt{2}\pi}V(\cT)$, at least for the on-shell tachyon.
Strictly speaking, the effective action $S_{DBI}$ \VTac\
is reliable only for the half S-branes. 
However, as we saw above, 
the disk partition function $Z=\bra B|0_{NS}\ket$
and the RR source $j=\bra B|C_{RR}\ket$ are correctly reproduced
by the effective action \VTac, \SCR\ even for the full brane.
It would be interesting to understand the physical meaning of this fact.

It is useful to rewrite $J$ as
\eqn\arctanJT{
J={1\o\pi}d\tan^{-1}\lf({T\o\rt{2}}\ri).
}
From this expression, it is obvious that the integral of $J$
is $1$ when integrated over
the kink configuration
of $T$ connecting the two vacua $T=\pm\infty$.
Therefore, $J=W(\cT)d\cT$ satisfies the condition \WvsVconsis.

The equation of motion of the 
bulk RR $p$-form field with the source term \RRcouple\ is given by
\eqn\Ceq{
(-\del_t^2+\del_{{\bf x}}^2) C_{1,\cdots,p}
=-j(t)\cob^{9-p}({\bf x}),
}
where ${\bf x}$ is the transverse coordinates of the non-BPS $Dp$-brane. 
The S-charge carried by the rolling tachyon is defined by \Sthermo
\eqn\Scharge{
Q_s(t)=\int d^{9-p}x F_{0,\cdots,p}=\int d^{9-p}x\,\del_t C_{1,\cdots,p}.
}
From this definition of S-charge,
one can show that $Q_s(t)$ is related to $j(t)$ 
in a simple way:
\eqn\delQt{
{d\o dt}Q_s(t)=j(t).
}
Using \arctanJT, we can integrate the relation \delQt\ to obtain
\eqn\Qsarc{
Q_s(t)={1\o\pi}\tan^{-1}\lf({T(t)\o\rt{2}}\ri),
}  
up to a constant shift.


We can check the general formula \delQt\ by the 
explicit computation as follows.
For the full brane,
the RR $p$-form generated by the source \Jtzero\
is 
\eqn\Cform{\eqalign{
C_{1,\cdots,p}=
{1\o \rt{2}\pi 2^\nu\Ga(\nu+1){\rm vol}(S^{8-p})}
\sum_{n=0}^{\infty}&(-1)^{n}\lf[
e^{{2n+1\o\rt{2}}(t-t_0)}-e^{-{2n+1\o\rt{2}}(t+t_0)}\ri]\cr
&\times\lf({2n+1\o\rt{2}r}\ri)^\nu K_\nu\lf({2n+1\o\rt{2}}r\ri),
}}
where $\nu={9-p\o2}-1$, ${\rm vol}(S^{d-1})={2\pi^{d/2}\o\Ga(d/2)}$,
and $K_\nu(z)$ is the modified Bessel function.
$r=\rt{{\bf x}^2}$ is the radial coordinate of the transverse directions. 
Plugging this expression into the definition \Scharge\ of $Q_s(t)$,
the S-charge of the full brane is obtained as
\eqn\Qsfull{
Q_{\rm full}(t)=Q_+(t)+Q_-(t)
}
where $Q_\pm(t)$ are the S-charge of the half S-branes:
\eqn\Qhalf{
Q_+(t)={1\o\pi}\tan^{-1}\Big(e^{{1\o\rt{2}}(t-t_0)}\Big),\quad
Q_-(t)={1\o\pi}\tan^{-1}\Big(e^{-{1\o\rt{2}}(t+t_0)}\Big).
}
It is easy to see that this expression satisfies 
\delQt.
As a function of time, $Q_{\rm full}(t)$ has a 
kink at $t=t_0$ and an anti-kink at $t=-t_0$, and  
approaches $\hf$ at large $|t|$. 
(More precisely, $Q_{\rm full}(t)$ has a ``half'' (anti)kink
at $t=\pm t_0$. See the discussion below.)
Note that the S-charge is independent of time when
$\la=\hf$, or $t_0=0$ \Sthermo 
\eqn\Qstzero{
Q_{\rm full}(t)\Big|_{t_0=0}=\hf.
}
This is consistent with the relation \delQt\ since 
there is no source of RR field when $\la=\hf$.

The S-charge $Q_s(t)$ at fixed $t$
is not a gauge invariant quantity.
The physical S-charge is the difference of $Q_s(t)$ in
the far future and the far past:
\eqn\cobQs{
\lap Q_s=Q_s(t=+\infty)-Q_s(t=-\infty).
}
This is the  definition of the integral of RR flux over the 
``transverse sphere'' surrounding S-brane in Lorentzian space \Sbrane.
For the half S-brane and the full brane, their physical S-charges are
\eqn\lapQpmf{
\lap Q_\pm=\pm\hf,\quad \lap Q_{\rm full}=0.
}
The S-charge $\lap Q_\pm=\pm\hf$ of half S-brane 
reflects the fact that tachyon field describing
the half S-brane runs from $T=0$ to $T=\infty$ 
which is the half way between the two vacua $T=\pm\infty$. 

The tachyon configuration of the form 
$\sinh t/\rt{2}$, which  connects the two vacua $T=\pm\infty$,
can be obtained from the $\cosh t/\rt{2} $ configuration \fullT\
by the shift $t\riya t+\pi i/\rt{2}$, $t_0\riya t_0+\pi i/\rt{2}$ \SenTM.
Under this shift of $t$ and $t_0$, $T$ in \fullT\ becomes
\eqn\shT{
{T\o\rt{2}}={\sinh{t\o\rt{2}}\o \cosh{t_0\o\rt{2}}}.
}
The on-shell Lagrangian (or the disk partition function) 
of this configuration
is the same as the cosh case \onL,
but the S-charge becomes
\eqn\Qsh{
Q_{\sinh}(t)=Q_+(t)-Q_-(t),
}
and it satisfies $\lap Q_{\sinh}=1$ (we normalized $j(t)$
by requiring this relation \Sthermo).

\newsec{Discussion}
In this paper, we showed that the source $j(t)$ of RR field
computed from the boundary state is reproduced by the
Wess-Zumino term of the form \WvsV.
We should comment on the relation between our $W(\cT)$ 
and the Wess-Zumino term obtained in BSFT \refs{\TakayaBSFT,\KrausBSFT}.
The function $W(\cT)$ in \WvsV\ is quite different from 
the BSFT result $W(\cT)\sim e^{-\cT^2/4}$.
As emphasized in \Kutasov, the two
actions have different regimes of validity, 
{\it i.e.} our $W(\cT)$ is valid for nearly on-shell configurations
while BSFT is valid far off mass shell.
However, in view of the topological nature of
WZ term in BSFT, namely the Chern character of superconnection,
we expect our $W(\cT)$ is related to that in BSFT by a nontrivial
field definition.

We have two pictures suggesting
there are some branes at $t=\pm t_0$ 
(or its shift in the imaginary direction):
\item{(1)} 
From the form of $Q_{\rm full}(t)$,
when $t_0$ is large 
the full brane can be well approximated by the following configuration:
The unstable $Dp$-brane exists only 
in the region $|t|<t_0$ and ends on the half $Sp$-branes
at $t=\pm t_0$, and there is no brane before $-t_0$ and after $t_0$.
It was discussed in \Drukker\ that 
the non-BPS branes in Euclidean space can end on
``half D-branes'' which carry the half unit of RR charge.
This is based on the fact that the unstable branes can be regarded as
sphalerons in string theory \Spha. 
\item{(2)}
In \refs{\LLM,\Sthermo}, it was observed that
when $t_0=0$ the closed string state produced by the 
unstable $Dp$-brane is closely related to the array of
BPS $D(p-1)$-branes at $t={\pi i}(2n+1)/\rt{2}$.
This comes from the fact that the disk partition function $Z(t)$
has poles at $t={\pi i}(2n+1)/\rt{2}$.
When we turn on $t_0$, the poles of $Z(t)$ and $j(t)$ are
shifted to $t=\pm t_0+{\pi i}(2n+1)/\rt{2}$.
The closed string emission rate from the full brane
is related to the cylinder amplitude between
(anti)D-branes sitting at $t=\pm t_0+\pi i(2n+1)/\rt{2}$ \LLM.

\noindent It would be interesting to study the relation between
these two pictures (see \Sthermo\ for a discussion
on the relation between S-branes and D-instantons). 

\vskip 1cm
\centerline{{\bf Acknowledgement}}
I would like to thank Ben Craps and
David Kutasov for useful comments and encouragement.

\listrefs
\bye